\documentclass[12pt]{article}
\usepackage{graphicx}
\usepackage{amssymb}

\newcommand{\be}{\begin{equation}}
\newcommand{\ee}{\end{equation}}
\newcommand{\ba}{\begin{eqnarray}}
\newcommand{\ea}{\end{eqnarray}}
\newcommand{\nn}{\nonumber \\}

\newcommand{\bfX}{{\bf X}}
\newcommand{\bfa}{{\bf a}}

\def\({\left (}
\def\){\right )}
\def\[{\left [}
\def\[{\right ]}

\def\t{\tilde}

\begin{document}
\begin{titlepage}
\bigskip
\rightline{}
\rightline{hep-th/0411268}
\rightline{NSF-KITP-04-128}
\bigskip\bigskip\bigskip\bigskip
\centerline {\Large \bf {How Hairy Can a Black Ring Be?}}
\bigskip\bigskip
\bigskip\bigskip

\centerline{\large Gary T. Horowitz$^1$ and Harvey S. Reall$^2$}
\bigskip\bigskip
\centerline{\em ${}^1$Department of Physics, UCSB, Santa Barbara, CA 93106}
\centerline{\em gary@physics.ucsb.edu}
\bigskip
\centerline{\em ${}^2$ Kavli Institute for Theoretical Physics, UCSB,
Santa Barbara, CA 93106}
\centerline{\em reall@kitp.ucsb.edu}

\bigskip\bigskip

\begin{abstract}
It has been shown recently that there is a large class of supersymmetric solutions of five-dimensional supergravity which generalize the supersymmetric black ring solution of Elvang et al.
This class involves arbitrary functions. We show that most of these solutions do not have smooth event horizons, so they do not provide examples of black objects with infinite amounts of ``hair".

\end{abstract}

\end{titlepage}

\baselineskip=18pt
\setcounter{equation}{0}

\section{Introduction}
In four dimensions, there is considerable evidence to support Wheeler's
idea that black holes have no hair. The traditional statement that
stationary black holes in Einstein-Maxwell theory are uniquely
characterized by their mass, charges and angular momentum was
established by the  uniqueness theorem for the
Kerr-Newman black hole. There are also results showing that
gravity coupled to other simple matter fields (e.g. scalar fields with suitable
potentials) does not yield any new stationary black hole solutions
\cite{Bekenstein:1996pn}.
Over the past decade, it
has been shown that more complicated matter fields can yield new
black holes. In particular, many theories with  gravitating solitons  have
new black hole solutions, since one can put a small black hole inside the
soliton \cite{Kastor:1992qy}.
However, even in these cases, it remains true that the black holes
are characterized by a small number of parameters. Most of the information that
falls into the black hole is (classically) lost. The spirit of the ``no-hair theorem" holds.

In more than four dimensions, much less is known. Even for pure gravity,
stationary five-dimensional black holes are not uniquely characterized by
their mass and angular momenta. There exist ``black rings" -- black holes of topology $S^1 \times S^2$ -- that
 have the same conserved quantities as black holes of spherical topology \cite{Emparan:2001wn}. If electromagnetic
  fields are included, then the black rings can carry charge
  \cite{Elvang:2003yy} and, moreover,
  the number of parameters required to specify black ring solutions now
 exceeds the number of conserved quantities that they carry 
 \cite{Emparan:2004wy}. This remains true even when one imposes the
strong constraint of supersymmetry. A family of $U(1)\times U(1)$ invariant
supersymmetric black rings has been found 
\cite{Elvang:2004rt, Bena:2004de, Elvang:2004ds,Gauntlett:2004qy} and these solutions
are not characterized by their asymptotic charges.
Nevertheless, for all of these examples, only a finite number of parameters are required to specify the solution so again the spirit of the no-hair theorem is retained.

Recently, a more serious challenge to the idea that black holes have no hair has arisen. Bena and Warner (BW) have claimed \cite{Bena:2004de} that there exists a family of supersymmetric black ring solutions described by {\it arbitrary functions}!

BW showed how to construct a family of supersymmetric solutions of
five-dimensional supergravity that is specified by seven arbitrary functions of
one variable. Four of these functions describe the embedding of a curve
$\cal C$ into $R^4$ and the other three specify certain charge densities
along $\cal C$. If $\cal C$ is taken to be a round circle with constant 
charge densities then this construction reproduces the $U(1)\times U(1)$
invariant black ring solutions. BW claimed that these restrictions on the functions are unnecessary and that black ring solutions exist for any smooth non-intersecting closed curve $\cal C$ with essentially arbitrary charge densities. They based this claim on the following argument. $\cal C$ can be approximated locally by a straight line with constant charge density. The solution in which $\cal C$ is exactly straight with constant charge density was constructed in \cite{Bena:2004wv} and describes an infinite black string
 with a smooth horizon. Hence the solution for arbitrary $\cal C$ with varying charge density should behave locally like that of \cite{Bena:2004wv} and therefore have a regular horizon.

We will show that this argument is incorrect: most of these
solutions do not have smooth horizons and hence cannot be regarded
as black ring solutions. In fact our arguments suggest that the
only solutions which are smooth belong to the original $U(1)\times U(1)$ 
invariant family 
(or superpositions of them \cite{Gauntlett:2004wh,Gauntlett:2004qy}). The situation
is similar to a straight extremal black string in five or six
dimensions wrapped around an $S^1$. Such solutions have a null
translational symmetry so one can add various types of traveling
waves described by arbitrary functions \cite{Garfinkle:1992zj}.
Soon after they were discovered, these waves were interpreted in
terms of ``classical hair" \cite{Larsen:1996ed}. However, it was
later shown that there is a problem with this interpretation.
 After the waves are
added, the metric is still continuous at the horizon and all scalar curvature
invariants remain finite \cite{Horowitz:1996th}. Nevertheless, it turns out
that there is a curvature singularity at the horizon because certain curvature
components in a basis parallelly propagated along a geodesic will diverge
there \cite{Horowitz:1997si,Kaloper:1996hr}. More physically, a freely falling
observer will experience infinite tidal forces. This is a result of the fact
that the waves travel around the circle an infinite number of times, accumulating near the horizon.

The problem with the BW solutions is similar. The BW argument assumes that a point on the curve $\cal C$ in $R^4$ corresponds to a point on a horizon in spacetime (more precisely, a point on the $S^1$ of a $S^1\times S^2$ horizon). However, we shall see that an actual horizon requires a limit  $\ell \rightarrow \infty$ where $\ell$ is the parameter along the curve.\footnote{This is because the solution rotates in the $\ell$ direction. This behaviour is familiar from the Kerr solution, for which the azimuthal angle of the Boyer-Lindquist coordinates diverges at the horizon.} So any non-trivial periodic function of $\ell$ (such as the charge densities on $\cal C$) will run through an infinite number of periods before crossing the horizon and hence will not be continuous there. We will use this to show that the metric does {\it not} admit a smooth horizon.

More precisely, we will show that the metric cannot be extended through a
$C^2$ horizon.  The similarity with the results of
\cite{Horowitz:1997si,Kaloper:1996hr} suggests that there will be a parallelly-propagated curvature singularity although we have not proved this. Although $C^2$ is the minimum degree of differentiability usually demanded of spacetime, one might ask whether the metric can be extended through a horizon of lower differentiability, such as the $C^0$ horizon of \cite{Horowitz:1996th}, as would be required to assign an area to the horizon. This remains an open question.\footnote{Note that the four dimensional uniqueness theorems assume smoothness (in fact analyticity) of the horizon so if one permits horizons of lower differentiability then it might be possible to violate black hole uniqueness even in four dimensions.}  In any case, the lack of a smooth horizon shows that these solutions do not really describe black holes with infinite hair of the type envisaged by Wheeler.

The $U(1)\times U(1)$ invariant black rings evade our argument because the charge densities are constant and $\cal C$ follows the orbit of a Killing field of $R^4$ so nothing depends on $\ell$. Hence the limit $\ell \rightarrow \infty$ can be taken and a smooth extension through a horizon found, as described in \cite{Elvang:2004rt,Elvang:2004ds,Gauntlett:2004qy}.

In the next section we review the equations that govern
supersymmetric solutions of five-dimensional supergravity and BW's
method of solving them. %%%%%%
We also clarify the (surprisingly subtle) relation between the
charge density on the ring and the total charge of the spacetime.
The difficulties of having a smooth horizon can be seen even in
the simple case when the curve ${\cal C}$ is a straight line, so
we discuss this first, in section 3. The case of an arbitrary
curve is investigated in section 4. The final section contains
some further discussion.

\setcounter{equation}{0}
\section{Supersymmetric solutions of 5D supergravity}
\label{sec:susy}

If we set all three charge densities equal, then the solutions of
BW become solutions to minimal five-dimensional supergravity. The
bosonic sector of this theory is just Einstein-Maxwell theory with
a Chern-Simons term. Any supersymmetric solution of this theory
admits a globally defined non-spacelike Killing vector field $V$
\cite{Gibbons:1993xt}. In a region where this is timelike, any
supersymmetric solution takes the form \cite{Gauntlett:2002nw}:
\ba
 ds^2 &=& -f^2 (dt + \omega)^2 + f^{-1} h_{mn} dx^m dx^n \nn
  F &=& \frac{\sqrt{3}}{2} d \left[ f (dt+ \omega) \right] - \frac{1}{\sqrt{3}} G^+
\ea
where $V=\partial/\partial t$,  $h_{mn}$ is a hyper-K\"ahler metric on a four-dimensional
``base space" $\cal B$, $f$ and $\omega$ are a scalar and 1-form on $\cal B$, and $G^+$ is defined by
\be
 \label{eqn:gplusdef}
 G^+ = \frac{1}{2} f \left( d\omega + \star d\omega \right),
\ee
where $\star$ denotes the Hodge dual with respect to $h_{mn}$ with orientation chosen so that the complex structures are anti-self-dual. The equations of motion for the theory reduce to the following equations on $\cal B$ \cite{Gauntlett:2002nw}
\be\label{Geq}
 dG^+ = 0,
\ee
\be
\label{eqn:maxwell}
 \qquad \Delta f^{-1} = \frac{4}{9} \left(G^+\right)^2.
\ee
where $\left(G^+\right)^2 \equiv {1\over 2} (G^+)_{mn} (G^+)^{mn}$ with
indices raised with the base space metric $h_{mn}$ and $\Delta$ is the Laplacian on $\cal B$.

A good understanding of these equations was lacking until Bena and Warner made the important discovery \cite{Bena:2004de} that they can be solved in a linear manner as follows. For simplicity, we take the base space to be flat: ${\cal B} = R^4- {\cal C}$ where $\cal C$ is a smooth curve without self-intersections. The first step is to solve (\ref{Geq}). Introduce a current $J$ on ${\cal C}$ and let $G$ solve the Maxwell equations for this current on $R^4$, i.e., $ d\star G = \star J$, $d G = 0$. Now set $G^+ = G + \star G$ on $R^4-{\cal C}$. This gives
\be
 dG^+ = \star J,
\ee
and hence (\ref{Geq}) is satisfied on $R^4 - {\cal C}$ because $J$ is supported on $\cal C$. 

Having determined $G^+$, the next step is to solve (\ref{eqn:maxwell}) to determine $f$. BW present an expression for the general solution using a Green's function integral. However, this integral does not converge because $G^+$ diverges as $1/d^2$ where $d$ is the proper distance from $\cal C$, so the right hand side of (\ref{eqn:maxwell}) diverges as $1/d^4$ and therefore is not integrable. Fortunately, this problem can be circumvented as follows.

Consider a Poisson equation
\be
\label{eqn:poisson}
 \nabla^2 \phi = s_0 + s_1
 \ee
where $s_0$ and $s_1$ are smooth functions on $R^4 - {\cal C}$ and $R^4$ respectively, and
$s_0+s_1$ vanishes sufficiently rapidly at infinity. Let $\cal T$ denote a neighborhood of $\cal C$. Then the solution satisfying $\phi=\bar{\phi}$ at infinity is given by $\phi = \phi_0 + \phi_1$ where $\phi_0$ obeys
\be
 \nabla^2 \phi_0 = s_0 \qquad \mbox{in ${\cal T}$},
\ee
and $\phi_0$ is chosen to decay smoothly to zero outside ${\cal T}$
(without satisfying any particular equation). $\phi_1$ is defined as the unique solution of
\be
 \nabla^2 \phi_1 = s_0 + s_1 - \nabla^2 \phi_0,
\ee
obeying $\phi_1=\bar{\phi}$ at infinity. The right hand side of this equation is smooth on $R^4$, so $\phi_1$ can be determined using a Green's function, and is smooth on $R^4$. Note that the singular part of $\phi$ is determined entirely by $\phi_0$. In other words, we can determine the singular part of $\phi$ just by solving the Poisson equation in the neighborhood of $\cal C$ without worrying about the boundary condition at infinity, as one would expect physically. Any two solutions of (\ref{eqn:poisson}) that obey the same boundary condition at infinity differ at most by a harmonic function on $R^4 - {\cal C}$ that vanishes at infinity. Such a function can be regarded as arising from a distributional source on $\cal C$.

Carrying out this procedure for equation (\ref{eqn:maxwell}) determines $f^{-1}$ up to the addition of a function harmonic on $R^4 - {\cal C}$ which can be regarded as arising from a charge distribution $\rho$ on $\cal C$ so, using differential forms, we write
\be
\label{eqn:rhodef}
 d \star d f^{-1} + \frac{4}{9} G^+ \wedge G^+ = 4\pi \rho \delta[{\cal C}] \star 1,
\ee
where $\rho$ is the charge density on $\cal C$ and $\delta[{\cal C}]$ is a delta-function supported on $\cal C$.

BW arrived at the surprising conclusion that the total charge of
the spacetime is not determined entirely by $\rho$ but also has a
contribution from fluxes. We shall see that this is incorrect. The
confusion arises from subtleties in the definition of $\rho$ so it
is worth discussing this at some length.

First, note that we have to move the singular $\left(G^+\right)^2$
term to the left hand side of (\ref{eqn:rhodef}) in order to
interpret the RHS as a distribution. In practice, we will use this
equation to determine $\rho$ once we have a solution for $f$ and
$G^+$. The solution for $f$ will involve an arbitrary coefficient
and (\ref{eqn:rhodef}) can be used to relate this coefficient to
$\rho$. The obvious way to do this is to integrate over a tubular
neighbourhood ${\cal T}$ of ${\cal C}$ to obtain \be
\label{eqn:rhocalc}
 4 \pi \int_{\cal C} \rho = \int_{\cal \partial {\cal T}} \left( \star df^{-1} + \frac{8}{9} C \wedge G^+ \right),
\ee
where $C$ is a potential for $G$ in ${\cal T}$, i.e., $G = dC$. The right hand side is sometimes called the Page charge. It is obviously gauge invariant under $C \rightarrow C + d\lambda$. However, if $\cal C$ is a closed loop then $\cal T$ has non-trivial first cohomology and hence there is the freedom to perform large gauge transformations $C \rightarrow C + \theta$ where $\theta$ is closed but not exact. Therefore the above expression is not gauge-invariant. See \cite{Marolf:2000cb} for further discussion of this issue. In the present case we can resolve this ambiguity as follows. First we can define a conserved charge by
\be
 {\bf Q} = \frac{1}{4\pi} \int_{S^3} \star d f^{-1}.
\ee
where the $S^3$ is a sphere at infinity in $R^4$. One can check that this is proportional to the usual definition of the electric charge as the integral of $\star_5 F$ over $S^3$ where $\star_5$ denotes the five-dimensional Hodge dual. Now choose a gauge in which $C$ is globally defined on $R^4 - {\cal C}$ and vanishes at infinity (for example, the gauge in which $C$ is given by the usual Green's function integral in $R^4$). Equation (\ref{eqn:rhodef}) can now be written\footnote{
Actually these equations are not the same but differ by a term proportional to $C \wedge \star J$ supported on $\cal C$. However, this term is not well-defined because $C$ is singular on $\cal C$. In practice we must adopt (\ref{eqn:rhodef2}) as our definition of $\rho$ since we need to be able to integrate the LHS in order to make sense of the delta function on the RHS.}
\be
\label{eqn:rhodef2}
 d \left( \star d f^{-1} + \frac{8}{9} C \wedge G^+ \right) = 4\pi \rho \delta[{\cal C}] \star 1.
\ee
Integrating this equation over $R^4$ we then obtain
\be
 {\bf Q}  = \int_{\cal C} \rho,
\ee
where we have used the fact that $C$ and $G^+$ vanish at infinity. Hence the total charge is determined by the integral of the charge density of the source, just as expected physically.

Note also that by integrating the LHS of (\ref{eqn:rhodef2}) over
$R^4 - {\cal T}$ we conclude that the RHS of  (\ref{eqn:rhocalc})
is gauge invariant (as it is just $4\pi {\bf Q}$) provided we
demand that $C$ be well-defined on $R^4 - {\cal C}$, not just on
${\cal T}$. This is possible because $R^4 - {\cal C}$ has
vanishing first cohomology so demanding that $C$ be globally defined 
excludes large gauge transformations.

The final subtlety concerns how one uses (\ref{eqn:rhodef2}) to calculate the local charge density $\rho$ (rather than the integral of $\rho$). Pick a point $p$ on $\cal C$. Choose some surface $S(p)$ that intersects $\cal C$ orthogonally at $p$. Let $n$ denote the unit normal to $S(p)$ and $B(p,\epsilon)$ a 3-ball in $S(p)$ centered on $p$ with radius $\epsilon$. If we contract (\ref{eqn:rhodef2}) with $n$ and integrate over $B(p,\epsilon)$ we obtain
\be
\label{eqn:rhocalc3}
 \rho(p) = \frac{1}{4\pi} \int_{B(p,\epsilon)} i_n dY,
\ee
where $i_n$ denotes the operation of contracting $n$ with the first index of a form, and
\be
 Y \equiv \star d f^{-1} + \frac{8}{9} C \wedge G^+.
\ee
A short calculation gives
\be
 (i_n dY)_{\mu \nu \rho} = n \cdot \nabla Y_{\mu\nu \rho} + \left[ d(i_n Y) \right]_{\mu\nu\rho}
  -3 K_{[\mu}{}^\sigma Y_{\nu\rho]\sigma} + \ldots
\ee
where $K_{\mu\nu}$ is the extrinsic curvature of $S(p)$ and the ellipsis denotes a term that vanishes when projected onto $S(p)$. Substituting into (\ref{eqn:rhocalc3}) we get
\be
 \rho(p) = \frac{1}{4\pi} \int_{\partial B(p,\epsilon)} i_n Y + \mbox{volume terms}.
\ee
If we now take $\epsilon \rightarrow 0$ then the volume contributions vanish (this can be checked explicitly using the solutions we shall present later) leaving
\be
\label{eqn:rhocalc2}
 \rho(p) = \frac{1}{4\pi} \lim_{\epsilon \rightarrow 0}  \int_{\partial B(p,\epsilon)} i_n \left( \star d f^{-1} + \frac{8}{9} C \wedge G^+ \right).
\ee

 As an example, let's calculate $\rho$ for the black ring
solution of \cite{Elvang:2004rt}. Following  the notation of that
paper, ${\cal C}$ is a circle of radius $R$. We take $C = (3q/4)
(1 + y) d\psi$. This satisfies the above requirement of being
well-defined on $R^4 - {\cal C}$ and vanishing at infinity. If the
point $p$ has coordinate $\psi$ then we define $S(p)$ to be a
surface of constant $\psi$ through $p$. We then choose
$B(p,\epsilon)$ to be the region $y < -1/\epsilon$ of $S(p)$.
Hence $\partial B(p,\epsilon)$ is an $S^2$ parametrized by
$(x,\phi)$. We have $n = (1/R)[1 + {\cal O}(\epsilon)]
\partial/\partial \psi$ and find that $Y_{\psi x \phi} = (Q/2) -
q^2 x + {\cal O}(\epsilon)$. Plugging these results into
(\ref{eqn:rhocalc2}) then gives $\rho = Q/(2R)$, i.e., the charge
density is constant and proportional to $Q$.
%%%%%%
There is no explicit contribution from the fluxes\footnote{ 
One manifestation of the confusion in \cite{Bena:2004de} is that, in their discussion of
 this solution (setting their three charges equal to obtain a solution of
minimal 5D supergravity), BW claim (after equation (42)) that the coefficient
of $x-y$ in their function $Z$ is proportional to the charge density
$\rho$. Our calculation using equation (\ref{eqn:rhocalc2}) reveals that this is incorrect.}.

Returning to the original task of solving the supergravity
equations, we still need to determine $\omega$. The latter is not
gauge invariant: a shift $t \rightarrow t'  = t - \lambda(x)$
gives $\omega \rightarrow \omega' = \omega + d\lambda$. It is
therefore convenient to work with the field strength $W =
d\omega$. From the definition of $G^+$ we see that $W$ must obey
\be \label{eqn:Weq} dW=0, \qquad W + \star W = 2f^{-1} G^+ \ee BW
showed how to solve these equations to determine $W$ up to the
addition of a closed, anti-self-dual two-form. In practice, one
can attempt to use this freedom to ensure that $\omega$ is
globally defined and that the resulting solution is free of closed
timelike curves.

\setcounter{equation}{0}
\section{Straight string}

In this section we shall consider a solution describing an infinite string,
in which the curve $\cal C$ is taken to be a straight line in flat space with
the charge density allowed to vary along the line. Such solutions were briefly
discussed in \cite{Bena:2004wv}, but not given explicitly. If the charge density is given by a periodic function then one can identify the solution to obtain a finite string wrapped on a Kaluza-Klein circle. We shall show that this solution has a smooth horizon only if the charge density is constant.

\subsection{The solution}

Take $\cal C$ to be a straight line $L$. Choose cylindrical polar coordinates centered on $L$:
\be
 h_{mn} dx^m dx^n = dz^2 + dr^2 + r^2 \left( d\theta^2 + \sin^2 \theta d\phi^2 \right).
\ee
Let $J$ be a constant density line source on $L$:
\be
 J  = -3\pi q dz \delta[L].
\ee
A solution for $C$ is
\be
 C = - \frac{3q}{4r} dz,
 \ee
which is well-defined on $R^4 - L$ and vanishes at infinity, as required.
The solution for $G^+$ is then
\be
 G^+ = -\frac{3q}{4 r^2} \left( dz \wedge dr + r^2 \sin \theta d\theta \wedge d\phi \right),
\ee
where we have chosen the orientation $\epsilon_{zr\theta \phi} = +1$. Equation (\ref{eqn:maxwell}) is
\be
 \nabla^2 f^{-1}  = \frac{q^2}{2  r^4},
 \ee
with solution
\be
 f(r,z)^{-1}  = \frac{q^2}{4  r^2} +
 \int^{\infty}_{-\infty} \frac{\rho(z') dz'}{ \pi \left( r^2 + (z-z')^2 \right)} + 1,
 \ee
where we have normalized $f = 1$ at $r=\infty$ and allowed for a varying charge density $\rho(z)$  on $L$.\footnote{
Equation (\ref{eqn:rhocalc2}) can be used to verify the $\rho$-dependence of this expression. Choose $S(p)$ to be the plane with the same $z$-coordinate as $p$ and $B(p,\epsilon)$ to be the ball $r<\epsilon$ on $S(p)$.}
Finally we can solve for $\omega$ starting from the ansatz
\be
 \omega = w(r,z) dz,
\ee
which leads to
\be
 \partial_r w =  \frac{3q^3}{8  r^4} + \frac{3q}{2 r^2} \int^{\infty}_{-\infty}
 \frac{\rho(z') dz'}{ \pi \left( r^2 + (z-z')^2 \right)} +
 \frac{3q}{2 r^2}.
 \ee
The integral can be computed if we assume that
\be
 \rho(z) =  {\rm Re} F(z)
\ee
where $F(z)$ is a holomorphic function with suitable behaviour as $|z| \rightarrow \infty$ with ${\rm Im} z >0$, and no poles in the upper half-plane (e.g. $F(z) = a + b \exp (iz)$):
\be
 \int^{\infty}_{-\infty} \frac{\rho(z') dz'}{ \pi \left( r^2 + (z-z')^2 \right)} = \frac{1}{r} {\rm Re} F(z+ir).
\ee
We can then expand in $r$ for small $r$:
 \be
 f^{-1} = \frac{q^2}{4 r^2} + \frac{1}{r} {\rm Re} F(z) + 1 -{\rm Im} F'(z) -
 \frac{r}{2} {\rm Re} F''(z) + {\cal O}(r^2)
 \ee
\be
 w = -\frac{q^3}{8 r^3} -\frac{3q }{4 r^2} {\rm Re}F(z) - \frac{3q }{2  r} \left( 1 - {\rm Im} F'(z) \right) - \frac{3q}{4} {\rm Re} F''(z) \log r + {\cal O}(r).
\ee
The $\log r$ term can be eliminated by a gauge transformation $t \rightarrow t'  = t - \lambda(x)$ at the expense of introducing a $r$-component into $\omega$.

\subsection{Geometry near the source}

We now examine the metric near $L$. We  have
\be
 g_{tt} = -f^2 = {\cal O}(r^4),
\ee
\be
 g_{tz} = -f^2 w = \frac{2r}{q} + {\cal O}(r^2),
\ee
\be
 g_{zz}  = f^{-1} - f^2 w^2 = \frac{p(z)}{q^2} + {\cal O}(r \log r),
\ee
where
\be
 p(z) \equiv 3 \left[ ({\rm Re}F(z))^2 + \left({\rm Im} F'(z) - 1\right)q^2 \right].
\ee
We shall assume that $F(z)$ is chosen so that $p(z)$ is positive.
The leading order metric near $r=0$ is
%We can define a near-horizon limit by setting $r = \epsilon \tilde{r}$,
%$t = \tilde{t}/\epsilon$ and taking $\epsilon \rightarrow 0$.
%Dropping the tildes, the resulting metric is
\be\label{nearhor}
 ds^2 = \frac{4r}{q} dt dz + \frac{p(z)}{q^2} dz^2 + \frac{q^2}{4} \frac{dr^2}{r^2} + \frac{q^2}{4} d\Omega^2.
\ee
On a constant $t$ surface, the horizon appears to be at $r=0$, and $z$
appears to be a good coordinate there. To see that this is incorrect,
consider first the special case $p={\rm constant}$. Using the symmetries,
one can easily check that all geodesics that approach $r=0$ have $|z|$ diverge.
To construct good coordinates, let
\be
  v = t- \frac{q \sqrt{p}}{4r}, \qquad \hat{z} = z - \frac{q^2}{2 \sqrt{p}} \log (r/q),
\ee
to get
\be
 ds^2 = \frac{4r}{q} dv d\hat{z} + \frac{2q}{\sqrt{p}} dv dr + \frac{p}{q^2} d\hat{z}^2 + \frac{q^2}{4} d\Omega^2,
\ee
which defines an analytic extension through a future horizon at $r=0$.
Note that $\hat{z}$ is a good coordinate on the horizon, which implies that
$z\rightarrow -\infty$ there.

Now return to the general case of varying $p$.
This metric is very similar to the one studied in \cite{Horowitz:1996th}.
As shown there, the metric is equivalent to $AdS_3\times S^2$.
Let $\sigma$ be a solution to
\be\label{sigmaeq}
 \sigma' (z)+ \sigma(z)^2  = \frac{p(z)}{q^4}
\ee
and set $\log H(z) = \int^z \sigma$. Introducing new coordinates
\be
 U = \int \frac{dz}{H(z)^2}, \qquad V ={2 t\over q^4} + \frac{ \sigma(z)}{2qr},
 \qquad Z = \sqrt{2qr} H(z),
 \ee
the metric (\ref{nearhor}) becomes
\be
 ds^2 =q^2\left [ Z^2 dU dV +  \frac{dZ^2}{Z^2}\right ] + \frac{q^2}{4} d\Omega^2,
 \ee
the product of an AdS$_3$ space of radius $q$ with an $S^2$ of radius $q/2$.
To understand where the horizon is  in these coordinates, we reconsider the case of constant $p$.
We choose $\sigma = -\sqrt{p}/q^2$. Then writing $(U,V,Z)$ in terms of
$(r,v,\hat{z})$ gives $H(z) = \exp(-z \sqrt{p} / q^2)$ and
\be
 U\propto r \exp\left( \frac{ 2\sqrt{p}}{q^2} \hat{z} \right),
  \qquad V \propto v , \qquad
  Z \propto \exp\left( - \frac{\sqrt{p}}{q^2} \hat{z} \right)
  \qquad {(\rm constant \ p)}
 \ee
The horizon is located at $r=0$ with finite $v$ and $\hat{z}$ and hence at $U=0$
with finite $V$ and (non-zero) $Z$. In the case of non-constant $p$, this implies that $H$, and hence $z$, must diverge at the horizon. Hence {\it a point on $L$ does not correspond to a point on the $AdS_3$ horizon.}

We have shown that the leading order metric (\ref{nearhor}) has a smooth
horizon at $r=0$. In fact, this horizon is completely homogeneous, even
when $p(z)$ is not constant. It does not retain any information about the varying charge distribution.
This horizon is also infinite in extent. However, we can periodically identify
the coordinate $z$ or, equivalently, take $\rho(z)$ to be a periodic function. The leading order metric then has a smooth finite  horizon.\footnote{This assumes that a periodic solution to (\ref{sigmaeq}) exists. An argument for this was given in \cite{Horowitz:1996th}.} However, this does not imply that the full solution has a  smooth horizon. The subleading terms in the metric  play  a crucial role. We will now show that these subleading terms behave so badly that the full metric does not admit a $C^2$ horizon.

 The metric has an $O(3)$ symmetry with $S^2$ orbits. Let ${\cal A}$ denote
the area of the $S^2$ through any point. If the metric had a $C^2$ (or even
$C^1$) horizon, then the first derivative of ${\cal A}$ off the horizon
would have to be continuous. But since
\be
 {\cal A} = 4 \pi r^2 f^{-1},
 \ee
we have
\be
  d {\cal A}|_{r=0} = 4\pi\rho(z) dr,
\ee
so $\rho(z)$ would have to be continuous at the horizon. But we have seen above that $z$ diverges at a horizon. Hence the existence of a horizon implies that $\rho$ must have a well-defined limit as $z$ diverges. This is impossible if $\rho(z)$ is a non-trivial periodic function. Hence the only solution with a compact $C^2$ horizon is the one with constant $\rho$.

There is a slight loophole in this argument since we have not shown that $dr$ is non-vanishing at the horizon when $\rho(z)$ is not constant. A more rigorous argument goes as follows. If the solution is $C^2$ then the connection is $C^1$ and the Killing spinor should be $C^2$. $f$ must also be $C^2$ because it is the square of the Killing spinor \cite{Gauntlett:2002nw}. From \cite{reall:03} we know that $f$ has a second order zero at the horizon for any solution with an $AdS_3 \times S^2$ near-horizon geometry. The explicit form of the solution is
\be
 f = \frac{4r^2}{q^2} + {\cal O}(r^3).
\ee
It follows that $r$ is $C^1$ with a first-order zero at the horizon. Hence $dr$ is $C^0$ and non-vanishing at the horizon.

\setcounter{equation}{0}
\section{Arbitrary curves}

In this section we consider the solution associated with an arbitrary closed curve in $R^4$.
We first introduce a convenient coordinate system, then find the solution near the horizon
(including the subleading contribution), and finally show that the horizon cannot be $C^2$.

\subsection{Fermi normal coordinates}

Consider a curve $\cal C$ in $R^4$ given by $x = x(\ell)$ where
$\ell$ denotes proper length along the curve (so $\dot{x}^2 = 1$).
We are primarily interested in a closed curve without
self-intersections so we shall assume that $\ell$ is a periodic
coordinate. Let $e_i(\ell)$ be three orthonormal vectors defined
along the curve that are normal to the curve (i.e. $\dot{x}(\ell)
\cdot e_i(\ell) =0$). Assign Fermi normal coordinates $(\ell,X^i)$
to the point a distance $|\bfX |$ along the straight line from $x
(\ell)$ in the direction $X^i e_i (\ell)$. These coordinates are
related to Cartesian coordinates by \be \label{fermi}
 x = x(\ell) + X^i e_i(\ell).
\ee
There is a natural choice of basis $e_i$ given by ($a,b,\ldots$ denote Cartesian coordinates)
\be
 e_i^a (\ell) = \dot{x}^b (\ell) J^{(i)}_b{}^a
\ee
where $J^{(i)}$ are the three anti-self-dual complex structures:
\be\label{J's}
 J^{(1)} = dx^1 \wedge dx^2 - dx^3 \wedge dx^4,
 \quad J^{(2)} = dx^1 \wedge dx^3 + dx^2 \wedge dx^4,
 \quad J^{(3)} = dx^1 \wedge dx^4 - dx^2 \wedge dx^3
\ee
This gives
\be
\label{eqn:fermiid}
 \dot{x} \cdot \dot{e}_i  (\ell) = - a_i (\ell), \qquad e_i(\ell) \cdot \dot{e}_j (\ell) = - \epsilon_{ijk} a_k (\ell), \qquad \dot{e}_i (\ell) \cdot \dot{e}_j (\ell) = \delta_{ij} \bfa(\ell)^2
\ee where $a \equiv \ddot{x}$ is the acceleration of the curve,
and $a_i (\ell) \equiv a(\ell) \cdot e_i (\ell)$. We shall regard
$a_i$ as the components of a 3-vector $\bfa$ in $R^3$ with
coordinates $X^i$. The second of these equations says that this
basis has rotation equal to minus the acceleration of the curve.
In Fermi coordinates, the flat metric on $R^4$ is \ba
 ds^2 &=& \left( 1 - 2 \bfa (\ell) \cdot \bfX + \bfa(\ell)^2 \bfX^2 \right) d\ell^2 + 2 \left( \bfa(\ell) \times \bfX \right)_i d\ell dX^i + \delta_{ij} dX^i dX^j, \nn
  &=& \left( 1 - \bfa(\ell) \cdot \bfX \right)^2 d\ell^2 + \left( d\bfX + \bfa(\ell)  \times \bfX d\ell \right)^2,
\ea
where $\bfa \cdot \bfX \equiv a_i X^i$ and $(\bfa \times \bfX)_i \equiv \epsilon_{ijk} a_j X^k$. Hence $\sqrt{h} = 1 - \bfa (\ell) \cdot \bfX$. These coordinates
are not valid globally but are well defined in a finite neighborhood
of the original curve. Note that (by construction) the surfaces of constant
$\ell$ are flat.
Everything above can be generalized to an arbitrary base space (the metric
then contains terms quadratic in $\bfX$ that involve the curvature).

As an example, consider a curve that is confined to the $12$-plane:
\be\label{curve}
 x^1 = x^1(\ell), \qquad x^2 = x^2 (\ell), \qquad x^3 = x^4=0.
\ee
Using (\ref{J's}), the basis vectors are
\be
 e_1 = ( - \dot{x}^2,\dot{x}^1,0,0), \qquad e_2 = (0,0,\dot{x}^1,\dot{x}^2), \qquad e_3 = (0,0,-\dot{x}^2,\dot{x}^1).
\ee
The Cartesian coordinates are related to the Fermi coordinates by
\ba
 x^1 &=& x^1(\ell) - X^1 \dot{x}^2 (\ell), \qquad x^2 = x^2(\ell) + X^1 \dot{x}^1(\ell) \nn
 x^3 &=& X^2 \dot{x}^1 (\ell) - X^3 \dot{x}^2(\ell), \qquad x^4 = X^2 \dot{x}^2(\ell) + X^3 \dot{x}^1 (\ell).
\ea
Note that the curve (\ref{curve}) is invariant under the flow of the vector field
\be
\label{eqn:kvf}
 x^3 \frac{\partial}{\partial x^4} - x^4 \frac{\partial}{\partial x^3} =
 X^2 \frac{\partial}{\partial X^3} - X^3 \frac{\partial}{\partial X^2},
\ee
and this vector field will consequently generate a symmetry of the resulting
solution, i.e., it will be a Killing vector field.
The components of the acceleration are
\be
 a_1 =  \dot {x}^1 \ddot{x}^2  - \dot{x}^2 \ddot{x}^1 , \qquad a_2 = a_3 = 0.
\ee
For a circle of radius $R$, $x^1=R\cos(\ell/R)$ and $x^2 = R\sin(\ell/R)$
so $a_1 = 1/R$ independent of $\ell$. If the charge density is constant then this solution
is the solution of \cite{Elvang:2004rt} with a regular $S^1 \times S^2$ horizon.
In this case, the coordinates $X^i$ are related to the ring coordinates $(\theta,\chi)$ regular on the $S^2$ of the horizon by $X^1 = r \cos \theta$,
$X^2 = r\sin \theta \cos \chi$, $X^3 = r \sin \theta \sin \chi$ hence $\bfX / |\bfX|$ is smooth on the $S^2$ of the horizon and the Killing field (\ref{eqn:kvf})
 is $\partial/\partial \chi$. In this example, $\partial/\partial \ell$ is also a Killing field.

\subsection{Solving the  equations}

Consider a source of constant density along the curve:
\be
 J = -3\pi q \dot{x} (\ell) \delta[\cal C].
\ee
Writing $G = dC$, the solution for $C$ in Cartesian coordinates is
\be
\label{eqn:Csol}
 C(x) =- \frac{3q}{4\pi} \int \frac{\dot{x}(\t\ell) d\t \ell}
 {\left[x - x(\t\ell) \right]^2},
\ee
and hence
\be
 G (x) = \frac{3q}{2\pi} \int \frac{ \left[ x - x(\t\ell) \right]
 \wedge \dot{x}(\t\ell) d\t\ell}{ \left[ x - x(\t\ell) \right]^4}.
\ee
To evaluate this integral, we first write the field point $x$ in
terms of Fermi normal coordinates $(\ell,X^i)$ using (\ref{fermi}) and expand
\be
 x(\t\ell) = x(\ell) + \dot{x}(\ell)(\t\ell -\ell) + {1\over 2}
\ddot{x}(\ell) (\t\ell -\ell)^2 + \ldots
\ee
Then
\be
\left[x - x(\t\ell) \right]^2 = \bfX^2 + (1-\bfa \cdot \bfX )(\t\ell -\ell)^2 + \ldots
\ee
Setting $\t\ell=\ell + |\bfX| \eta$, one can expand the integral for small
$|\bfX|$ to obtain
\be
 G(\ell,\bfX)  = -\frac{3q}{4  |\bfX|^3 } \dot{x} (\ell,\bfX) \wedge e_i(\ell,\bfX) \left( X^i +
 \frac{1}{2} X^i X^j  a_j (\ell) + \frac{1}{2} \bfX^2 a_i(\ell) \right) + {\cal O}(1)
\ee
In this expression, the orthonormal basis $(\dot x(\ell), e_i(\ell))$ defined along
the curve is extended to a global orthonormal basis $(\dot x(\ell,\bfX),e_i(\ell,\bfX))$ by
translating along the (flat) surfaces of constant $\ell$. Now we need to rewrite this basis in the Fermi coordinates. Using equations (\ref{fermi}) and (\ref{eqn:fermiid}) we obtain
\be
 \dot{x}(\ell,\bfX) = \frac{1}{1- \bfa \cdot \bfX} \left[ \frac{\partial}{\partial \ell} - \left( \bfa \times \bfX \right)_i \frac{\partial}{\partial X^i} \right], \qquad e_i(\ell, \bfX) = \frac{\partial}{\partial X^i},
\ee
or, in terms of covectors:
\be
 \dot{x}(\ell,\bfX) = \left( 1 - \bfa \cdot \bfX \right) d\ell, \qquad e_i (\ell,\bfX) = dX^i + \left(\bfa \times \bfX \right)_i d\ell.
\ee
Hence we have
\be
G= -\frac{3q}{4  |\bfX|^3 } d\ell \wedge \left( X^i - \frac{1}{2} X^j a_j (\ell) X^i + \frac{1}{2} \bfX^2 a_i (\ell) \right) dX^i + {\cal O}(1).
\ee
Choosing orientation $\epsilon_{\ell ijk} = \epsilon_{ijk}$ we then get
\ba
 G^+ &=&  - \frac{3q}{4  |\bfX|^3 }  \left[ \left( X^i - \frac{3}{2} X^j a_j (\ell) X^i + \frac{3}{2} \bfX^2 a_i (\ell) \right) d\ell \wedge dX^i \right. \nn
 & +& \left.  \frac{1}{2} \epsilon_{ijk} \left( X^k + \frac{1}{2} X^j a_j (\ell) X^k + \frac{1}{2} \bfX^2 a_k (\ell) \right) dX^i \wedge dX^j  \right] + {\cal O}(1).
\ea
Note that the leading terms are independent of the acceleration of the
curve and agree with the solution for a straight string.

Now we determine $f$ by solving
\be
 \Delta f^{-1} =  \frac{4}{9} (G^+)^2 =   \frac{q^2}{2  |\bfX|^4} (1 + 2\bfa \cdot \bfX ) + {\cal O}(1/\bfX^2).
\ee
The boundary condition is that $f \rightarrow 1$ at infinity in $R^4$ but this is not required to extract the singular small $\bf X$ behavior of $f^{-1}$ which can be done using the method of section \ref{sec:susy}. The result is
\be
 f^{-1}  = \frac{q^2}{4  \bfX^2 } ( 1 - \bfa(\ell)  \cdot \bfX) + \frac{b(\ell)}{ |\bfX| } + {\cal O}(1).
\ee
The $1/|\bfX|$ term arises from the freedom to add a harmonic function to $f^{-1}$ so that (\ref{eqn:rhodef2}) is satisfied with charge density $\rho(\ell)$. The explicit relationship between $b$ and $\rho$ is easy to obtain from equation (\ref{eqn:rhocalc2}) for the case in which $\cal C$ is a circle of radius $R$ because we already know the full solution for $C$ and $G^+$ in that case (see the discussion below equation (\ref{eqn:rhocalc2})). We find that
 \be
 \label{eqn:brho}
 b(\ell) = \rho(\ell) - q^2/(2R) \qquad \mbox{(${\cal C}$ a circle of radius $R$)}.
\ee
Unfortunately, we have not been able to obtain a simple result expressing $b(\ell)$ in terms of $\rho(\ell)$ for a general curve $\cal C$. The problem is that   in order to compute $\rho$ using (\ref{eqn:rhocalc2}) it is necessary to expand the integral (\ref{eqn:Csol}) for $C$ to second non-trivial order in $\bfX$. In particular, we need to know the $\bfX$-independent part of the $\ell$ component of $C$. There does not appear to be a local expression for this term for a general curve $\cal C$ so all we can conclude is that $b(\ell) = \rho(\ell) - c(\ell)$ for some function $c$ that depends only on the curve $\cal C$.

The final step is to calculate $\omega$. The leading order solution for
$\omega$ can be obtained from the straight string. The next order
is ${\cal O}(1/\bfX^2)$, where we will find terms linear in $\bfa$
or $b$. The $b$-dependent correction is independent of
$\bfa$ and hence should be the same as for the straight string. To
calculate the $\bfa$ dependent corrections, we note that the only
possible ${\cal O}(1/\bfX^2)$ term in $\omega_\ell$ is
proportional to $\bfa \cdot \bfX / |\bfX|^3$. In $\omega_i$ there
are three possibilities, proportional to $(\bfa \times \bfX)_i
/|\bfX|^3$, $a_i/ \bfX^2$ and $\bfa \cdot \bfX X^i /|\bfX|^4$
respectively. Plugging a general linear combination of these terms into equation (\ref{eqn:gplusdef}),
we find that the final two terms appear with coefficients such
that they combine into a total derivative, which can be eliminated by 
shifting in the time coordinate $t$. This leaves 
\be
 \omega = -\frac{q^3}{8 |\bfX|^3} \left[ \left( 1 - \frac{3}{2} \lambda \bfa \cdot \bfX \right) d\ell -
 \frac{3}{2} (1-\lambda) (\bfa \times \bfX)_i dX^i \right] - \frac{3q b}{4  \bfX^2} d\ell + {\cal O}(1/|\bf X|),
\ee
where $\lambda$ is an arbitrary constant related to the freedom to add a closed anti-self-dual term to $d\omega$. We determine $\lambda$ by noting that the $\ell i$ and $\ell\ell$ components of the five-dimensional metric diverge at $\bfX = 0$ unless we choose $\lambda = 5/3$. Therefore we must make this choice to ensure that $\cal C$ has finite proper circumference.

\subsection{Absence of a $C^2$ horizon}

If the above solution admits a smooth horizon then a result of
\cite{reall:03} establishes that the near-horizon geometry must be
$AdS_3 \times S^2$. We have seen already how this geometry arises
as the leading order metric of a straight string. Now we shall see
how it arises for a general curve $\cal C$. First we need to be
more precise about what we mean by ``leading order  metric".
Obviously this must involve a limit $\bfX \rightarrow 0$ so we
start by rescaling $\bfX \rightarrow \epsilon \bfX$. To keep the
metric non-degenerate as we send $\epsilon \rightarrow 0$ we must
also rescale $t \rightarrow t/\epsilon$ (so $t$ diverges as
$\epsilon \rightarrow 0$, as expected at a horizon). The resulting
metric is \be \label{nearhor2}
 ds^2 = \frac{4r}{q} dt d\ell + \frac{p(\ell,\theta,\chi)}{q^2} d\ell^2 + \frac{q^2}{4} \frac{dr^2}{r^2} + \frac{q^2}{4} \left( d\theta^2 + \sin^2 \theta d\chi^2 \right).
\ee
where we have parametrized $\bfX$ with spherical polar coordinates $(r,\theta,\chi)$ and $p(\ell,\theta,\chi)$ is a function that cannot be determined just from the leading order form of the solution presented above -- one would have to go to third order in $|\bfX|$ to calculate it, just as for the straight string. However, it turns out that this does not matter. Since we know that the above metric has to be $AdS_3 \times S^2$, it is easy to show that $p$ must be independent of $\theta$ and $\chi$.\footnote{By calculating the eigenvalues of $R^\mu_\nu$ one can deduce that $\partial/\partial \theta$ and $\partial /\partial \chi$ must be tangent to the $S^2$ of $AdS_3 \times S^2$. It then follows from the above metric that $\partial/\partial t$, $\partial /\partial \ell$ and $\partial /\partial r$ must be tangent to $AdS_3$. This implies that the Riemann tensor must factorize in the above coordinate system. This implies $\partial_\theta p = \partial_\chi p = 0$.}
The above metric is then the same as (\ref{nearhor}) (with $z \rightarrow \ell$) and hence manifestly $AdS_3 \times S^2$ with a smooth horizon at $r=0$ at which $\ell$ diverges. We can now use this fact to deduce that the full metric does {\it not} have a smooth horizon in general.

In order to argue that the solution does not admit a $C^2$ horizon, we shall repeat the strategy used for the straight string. We shall assume that there {\it is} a $C^2$ horizon and arrive at a contradiction by constructing a scalar whose gradient is not continuous at the horizon. Since the solution is no longer spherically symmetric, we cannot use the area of two-spheres as we did for the straight string. Instead, consider
\be
 f^2 \left(G^+ \right)^2 = \frac{18}{q^2} \left( 1 + 4 \bfa(\ell) \cdot \bfX - \frac{8b (\ell)}{q^2} | \bfX | \right) + {\cal O}(\bfX^2).
\ee
The left hand side is a scalar invariant of the solution. The analysis of \cite{reall:03} reveals that it should be at least $C^1$ at the horizon.  Writing $\bfX = r \hat \bfX$, we obtain
\be\label{deriv}
d \left[f^2 \left(G^+ \right)^2\right] =
\frac{72}{q^2}\left(\bfa (\ell) \cdot \hat\bfX -\frac{2b(\ell)}{q^2}\right) dr
\ee
at $r=0$. The left hand side should be continuous at the horizon. As before, we can now use the fact that, if the solution is $C^2$, then $f$ is $C^2$ with a second order zero on the horizon. This implies that $r \equiv |\bfX|$ is $C^1$ with a first order zero at the horizon so $dr$ is continuous and non-vanishing there. From the leading order metric, we know that $\theta$ and $\chi$ are continuous at the horizon, and hence so is $\hat \bfX$. Since $\ell$ diverges at the horizon, the only way that the right hand side can be continuous is if the periodic functions $\bfa$ and $b$ are independent of $\ell$. Existence of a $C^2$ horizon therefore implies that $\bfa$ and $b$ must be independent of $\ell$. Note that $b(\ell)$  cannot cancel the acceleration dependence since the acceleration term is direction dependent.

We have concluded that existence of a $C^2$ horizon requires that $b$ and $\bfa$ be independent of $\ell$. The latter requirement is equivalent to
\be
 0= \frac{da_i}{d\ell}= \frac{d^3{x}^a}{d\ell^3} \frac{dx^b}{d\ell} J^{(i)}_{ab}.
\ee
If a two-form has vanishing contraction with
all of the $J^{(i)}$ then it must be self-dual. Any non-zero self-dual
form has rank four. However, the two-form above clearly has rank two.
It must therefore vanish. Hence we must have
\be\label{strongcond}
 \frac{d^3{x}^a}{d\ell^3} \propto \frac{dx^a}{d\ell}.
\ee
The coefficient of proportionality probably has to be constant since otherwise we can construct a $\ell$-dependent scalar by taking the dot product of the LHS and RHS, and this scalar will probably appear in the expansion of the solution at next order and lead to a lack of smoothness at the horizon. Hence, after a shift in $x^a$, we get
\be
  \frac{d^2{x}^a}{d\ell^2} \pm \frac{x^a}{R^2} = 0,
\ee
for some constant $R$, with solutions given by straight lines, circles or hyperbolae. However, the latter do not satisfy the normalization condition $\dot{x}^2 = 1$ so must be excluded. The circles always lie in a 2-plane. 

Note that it is not sufficient for the curve to follow a symmetry of the
base space metric. For flat $R^4$, a helical curve has a tangent
vector which is a linear combination of a rotation and a translation, and
hence is a Killing field. Nevertheless, this curve does not satisfy
(\ref{strongcond}). The problem is that even though $\ddot x$ is not
affected by the added translation, the basis vectors $e_i$ are, so the
components of the acceleration $a_i$ pick up extra $\ell$ dependence.

%%%%%%%
To summarize, modulo a plausible assumption about the coefficient
of proportionality in (\ref{strongcond}), we have shown that the
only asymptotically flat solutions with a $C^2$ horizon are those for which $\cal C$
is a circle and $b(\ell)$ is constant. Since we know the precise
relationship between $b$ and the charge density $\rho$ for
circular $\cal C$ (equation (\ref{eqn:brho})), we can conclude
that the charge density $\rho$ is constant. So it seems that the only solution with a
$C^2$ horizon is the black ring of \cite{Elvang:2004rt} (and in that case, the horizon is actually smooth). 

\section{Discussion}

Recently, Bena, Wang and Warner have constructed some explicit solutions describing
a circular ring with three varying charge densities \cite{Bena:2004td}.
They show that one can adjust the charge densities
so that the function $p(\ell)$ in the near horizon metric (\ref{nearhor2}) is constant.
Since this is only one constraint on three functions, there are still two free functions worth of freedom.
However we have seen that a constant $p(\ell)$ is not necessary for the
leading order horizon metric to be smooth, and not sufficient to ensure
that the full metric  is smooth at the horizon. Even though their metrics are not solutions to minimal five-dimensional supergravity, the argument given in the previous
section is easily extended to show that their solutions do not have smooth horizons.
Their solutions contain extra scalars which can be viewed as the radii
of the internal circles in a toroidal compactification of M-theory.
Derivatives of these scalars off the horizon depend on $\ell$ and
will not be continuous on the horizon.

We have focussed on the question of smoothness of the horizon
since this is crucial for the no hair theorems in classical
general relativity. The fact that most of the solutions described
in \cite{Bena:2004de} do not have smooth horizons does not mean
that they have no physical interest. In the analogous situation of
traveling waves along extremal black strings, even though the
curvature diverges at the horizon, the metric is continuous there
so the horizon area is well defined. It turns out that this area
depends on the wave profile and one can reproduce the entropy
of the black string by counting string states with the prescribed
wave profile \cite{Horowitz:1996th}. It would be interesting to
know if the metric of the BW solutions is also continuous at
the horizon. One might think that this is obvious since we have
shown that the leading order metric near the horizon is locally
$AdS_3\times S^2$. However this only shows that the leading order
divergences in the metric are purely coordinate effects. Any
subleading divergences still remain to be addressed. The entropy
of the $U(1)^2$ invariant supersymmetric black rings of \cite{Elvang:2004rt, Bena:2004de, Elvang:2004ds,Gauntlett:2004qy} has recently been reproduced by a
microscopic counting of states in M theory \cite{Andy} (see
\cite{Bena:2004tk} for an earlier discussion of this problem). If
the more general solutions turn out to have a continuous horizon
with well defined area,  one could try to extend this argument to
 reproduce their
entropy.

\vskip 1cm \centerline{\bf Acknowledgments} \vskip .5cm

It is a pleasure to thank H. Elvang, R. Emparan, D. Marolf and D. Mateos 
for discussions. This paper was finished while G.H. was visiting the IAS in Princeton and he
thanks them for their hospitality. This work was supported in part
by NSF grants PHY-0244764 and PHY99-07949.

 \end{document}